\documentclass[useAMS,usenatbib]{mn2e}
\usepackage{lscape}
\usepackage{amsmath,amssymb}
\usepackage{graphicx,mathptm} 
\usepackage{times} 
\usepackage{natbib}
\usepackage{psfig}
\usepackage{aas_macros}
\usepackage{xcolor}

\def\apjl{Astrophys. J. Lett.}
\def\apjs{Astrophys. J. Suppl.}

\def\aap{Astron. Astrophys.}

\title[H(z) with cosmic chronometers at z$\sim$2] {Raising the bar: new constraints on the Hubble parameter with cosmic chronometers at z$\sim$2}

\author[Moresco] {Michele Moresco$^{1}$\thanks{E-mail: michele.moresco@unibo.it}\\ 
  $^1$Dipartimento di Fisica e
  Astronomia, Universit\`a di Bologna, Viale Berti Pichat 6/2, I-40127
  Bologna, Italy
}

\begin{document}

\maketitle

\begin{abstract}
One of the most compelling tasks of modern cosmology is to constrain the expansion history of the Universe, since this
measurement can give insights on the nature of dark energy and help to estimate cosmological parameters.
In this letter are presented two new measurements of the Hubble parameter H(z) obtained with the cosmic chronometer 
method up to $z\sim2$. 
Taking advantage of near-infrared spectroscopy of the few very massive and passive galaxies observed at $z>1.4$ 
available in literature, the differential evolution of this population is estimated and calibrated with different
stellar population synthesis models to constrain H(z), including in the final error budget all possible sources of systematic 
uncertainties (star formation history, stellar metallicity, model dependencies). This analysis is able to extend significantly 
the redshift range coverage with respect to present-day constraints, crossing for the first time the limit at $z\sim1.75$.
The new H(z) data are used to estimate the gain in accuracy on cosmological parameters with respect to previous measurements
in two cosmological models, finding a small but detectable improvement ($\sim$5 \%) in particular on $\Omega_{M}$ and $w_{0}$. 
Finally, a simulation of a Euclid-like survey has been performed to forecast the
expected improvement with future data. The provided constraints have been obtained just with the cosmic chronometers approach,
without any additional data, and the results show the high potentiality of this method to constrain the expansion
history of the Universe at these redshifts.
\end{abstract}

\begin{keywords} 
cosmology: observations, dark matter, dark energy
\end{keywords}


\section {Introduction}
\label{sec:intro}

Since the pioneering works of \citet{riess1998} and \citet{perlmutter1999}, the comprehension of the accelerated expansion of the Universe has become a key
question in cosmology. Amongst many proposed approaches to tackle this open issue \citep[for a comprehensive review, see][]{weinberg2013}, 
the most common ones rely on probes which are able to disentangle the evolution with redshift due to the expansion 
of the Universe, and the intrinsic evolution of the considered probe; for this purpose, generally probes with ``standard'' properties have been considered,
such as {\it standard candles} (Supernovae Type Ia, SNe) and {\it standard rulers} (Baryon Acoustic Oscillations, BAO). 
In this context, each analysis has its own strength and weakness, but it is now common understanding that it is of primary importance to explore various possibility
at the same time, to keep under control the systematic effects biasing each single probe, and to cut down the errors on cosmological parameters by combining the 
different probes.

As suggested by \citet{jimenez2002}, an alternative method to constrain the expansion history of the Universe is based on the study of the redshift evolution of
``cosmic chronometers''. In this approach, the relative age of passive galaxies $dz/dt$ can be used as {\it standard clocks}, since $H(z)=-1/(1+z)dz/dt$.
The key issues of this method are the identification of a proper population of galaxies to trace homogeneously the differential age evolution of the Universe, and
the reliance on theoretical models to estimate galaxy ages, and hence the degeneracy between parameters. As pointed out by many authors \citep[for an extensive 
review, see][]{renzini2006}, massive and passive early-type galaxies represent the ideal population to address the first point, having assembled their mass at high 
redshift and being passively evolving since then. 
To solve the second point, in previous papers \citep{moresco2011,moresco2012a} an improvement of the technique was proposed based on the study of a direct 
observable of galaxy spectra, the 4000~\AA~ break (D4000), which is known to be strictly correlated with galaxy age and metallicity, and less significantly dependent
on star formation history (SFH); since the assumption of a linear relation between $D4000$ and age of a galaxy has been proven to be an extremely good 
approximation for most $D4000$ regimes \citep[see][]{moresco2011,moresco2012a}, i.e. $D4000=A(SFH, Z/Z_{\odot})\cdot {\rm age} + B$, it is possible to redefine H(z) as:
\begin{equation}
H(z)=-\frac{1}{1+z} A(SFH, Z/Z_{\odot}) \frac{dz}{dD4000}
\label{eq:HzD4000}
\end{equation}
The strength of this new equation is that now the dependence on statistical and systematic effects has been decoupled: the factor $dz/dD4000$ is only dependent 
on observables, while all the dependence on degeneracy between parameters or assumption of stellar population synthesis (SPS) models is contained in the factor 
$A(SFH, Z/Z_{\odot})$, which can be calibrated on SPS models. Its drawback is that it relies on the estimate of the stellar metallicity of the population, or at least 
on having a reasonable proxy of it, to be used as a prior. All the possible systematic error deriving from these assumptions (both on models and on the galaxy properties)
have been considered and taken in consideration in the total error budget, as discussed in the following section.
In \cite{moresco2012a} was demonstrated that this new approach relying on the spectroscopic differential evolution of cosmic chronometers is robust against 
the choice of SPS models, and not significantly dependent on the SFH assumptions (also because the selection criteria leaves small space for prolonged SFHs). In that 
paper, eight new H(z) measurements have been provided in the redshift range $0.15<z<1.1$, significantly extending the redshift coverage and precision of previous similar 
analysis \citep{simon2005,stern2010}; the potential of this new method in comparison with more standard probes has been studied by many authors 
\citep[e.g. see][]{jimenez2012,moresco2012b, zhao2012,wang2012,sorensen2013,verde2014}, which demonstrated how the cosmic chronometers method can 
be competitive for many aspects with SNe and BAO in constraining cosmological parameters.

In this letter two new $H(z)$ measurements up to $z\sim2$ are reported, taking advantage of various literature data of high redshift ($z>1.4$) massive and passive galaxies with 
near-infrared spectroscopy covering the restframe range around the D4000. These data, never used before for this scope, represent a golden mine to extend present-day 
estimates of $H(z)$ to a redshift range never approached before with this technique, and matched only by the measurement of the BAO in the Ly$\alpha$ forest of BOSS 
quasars by \cite{busca2013}. Throughout this letter the new data are presented, and the gain in accuracy for different cosmological parameters discussed, providing also forecasts 
of improvement which may be reached with future surveys.


\section {Data}
\label{sec:data}

Different literature data providing spectroscopical analysis of high redshift ($z>1.4$) massive and passive galaxies are considered. The most massive, passive and 
with the shortest SFH sample has been selected from each analysis: these galaxies represent the most appropriate probe to sample the expansion history of the Universe,
mapping the ``eldest crust'' population at each redshift independently of the selection criteria. 
To apply the method, it has been considered the definition of the break with narrower bands ($D4000_{n},  $3850-3950 and 4000-4100 \AA) introduced by 
\cite{balogh1999}, being less sensitive to potential reddening effects. While some works provide such measurement, for the others the spectra have been analyzed, obtaining 
for the first time $\rm D4000_{n}$ estimates for these samples.

\begin{figure}
\includegraphics[angle=-90, width=0.47\textwidth]{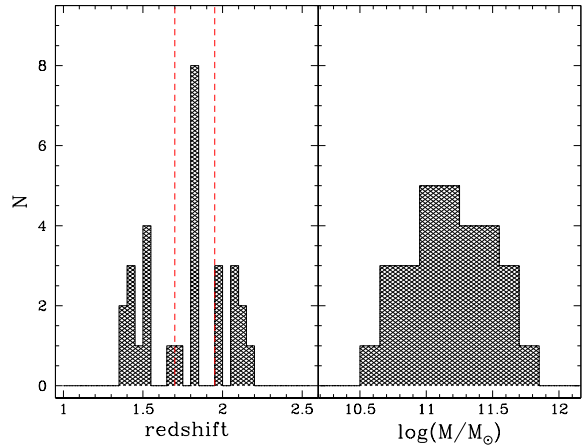}
\caption{Redshift (left plot) and stellar mass (right plot) histograms for the high-z passive galaxies analyzed. The red dotted lines show how the sample has been
divided into three redshift bins.}
\label{fig:hist_massz}
\end{figure}

\begin{figure*}
\includegraphics[angle=-90, width=0.89\textwidth]{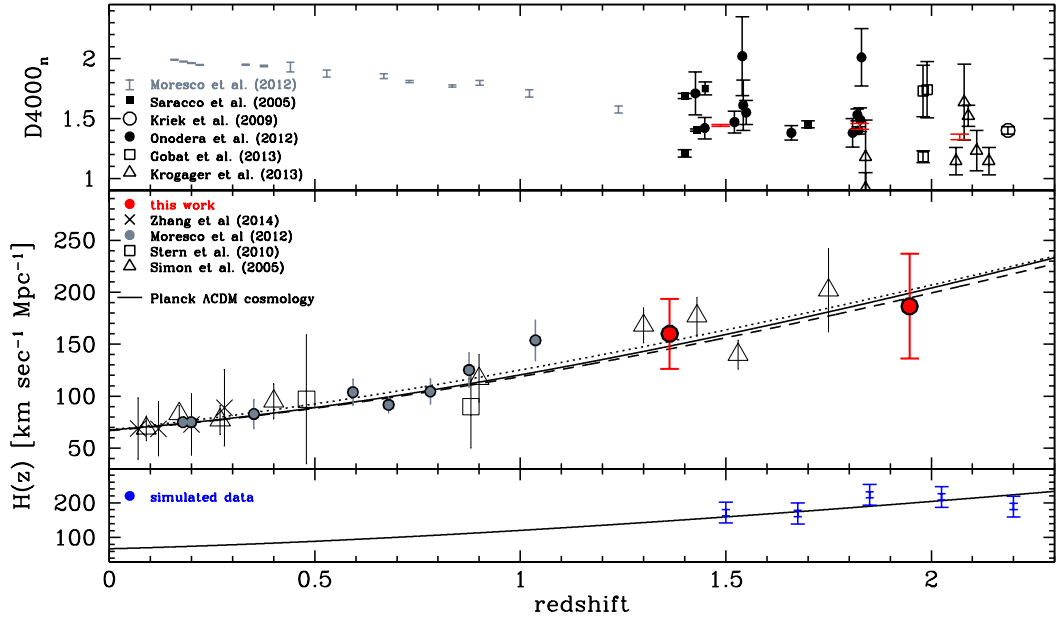}
\caption{Upper plot: D4000$_{n}$-z relation for the galaxies considered in this analysis (black points), their weighted average (red points) and the median
relation obtained at lower redshifts in \citet{moresco2012a} in the same mass regime. Intermediate plot: Hubble parameter measurements obtained (red points), 
compared with literature data. Filled points have been obtained with D4000-based cosmic chronometers, while the open points from the relative age based method.
The theoretical relation is not a best-fit to the data, but the $H(z)$ obtained assuming a Planck fiducial cosmology 
\citep[$\Omega_{m}=0.315$, H$_{0}$=67.3 km/sec/Mpc,][]{planck}. The best fits to the data for a open $\Lambda$CDM cosmology and for a flat wCDM
cosmology are shown respectively with a dotted and dashed line. Lower plot:  Simulated H(z) data from a Euclid-like survey. The inner errorbar represent the statistical
error, while the outer errorbar the total error.}
\label{fig:Hz}
\end{figure*}

In the following, the samples used, and how they have been selected, are briefly presented. For the full description these analyses, one can refer to the original papers.
\begin{itemize}
\item \citet{saracco2005}. This work presents seven massive evolved galaxies spectroscopically identified in the MUNICS survey; 
the spectral analysis confirmed their passive nature, with short SFH ($\tau<0.3$ Gyr) and high masses ($M/M_{\odot}>10^{11}$).
\item \citet{kriek2009}. They analyze the spectrum of a quiescent, ultra-dense galaxy at z=2.1865 obtained with the Gemini Near-Infrared 
Spectrograph. The spectral fitting provided a low star formation rate (present SFR$<$1\% of past SF), and large mass ($M/M_{\odot}>10^{11}$).
\item \citet{onodera2012}. In this work, 18 passive BzK galaxies in the COSMOS field at $z>1.4$ were spectroscopically observed 
and analyzed; the fit to their SED provided stellar masses $M/M_{\odot}=4--40\times10^{10}$. The most passive sample is further selected, 
discarding galaxies with 24 $\mu$m detection, prolonged  SFH ($\tau>=0.3$ Gyr) and high H$\delta$ values (EW(H$\delta)<$-3\AA).
\item \citet{gobat2013}. From HST/WFC3 slitless spectroscopic observations of the cluster Cl J1449+0856, 6 quiescent members were identified at $z=2$;
four of them have a high enough S/N to allow the measurement of D4000, and their stellar masses were measured in \citet{strazzullo2013} 
($M/M_{\odot}>1.2\times10^{11}$).
\item \citet{krogager2014}. In this work, they presented deep HST/WFC3 grism spectroscopy for a sample of 14 galaxies at $z\sim2$ in the COSMOS field.
Among them, 7 are quiescent galaxies with measured D4000 and masses $M/M_{\odot}>0.6\times10^{11}$. Due to the low resolution and poor sampling
of the grism data, the D4000 has been measured in its definition with larger bands \citep[$D4000_{w}$, 3850-3950 and 4000-4100 \AA,][]{bruzual1983}, and to compare
it with the other data SPS models are used to obtain the required conversion\footnote{Note that the conversion factor from $D4000_{w}$ to $D4000_{n}$ in this range of 
values is $\sim$12\%, and does not significantly depend on the assumed model}.
\end{itemize}
In this way, 29 galaxies are selected, which represent the most massive, red and passive galaxies found so far at these redshifts. On average, the formation 
redshifts (estimated from the best-fit to both galaxy spectra and photometry) are $3.5<z_{f}<5$, with a spread in formation age $<0.45$ Gyr, which is much smaller than the associated error, 
therefore avoiding possible progenitor bias issues.
In Fig. \ref{fig:hist_massz} are shown the redshift and stellar mass distributions of the final sample. 
The scatter in $D4000_{n}$ between the analyzed galaxies ($\sigma_{D4000}\sim$0.2-0.25) is statistically consistent with the properties of the considered population, 
with the $D4000_{n}$ errors and the estimated SFH contributing in similar fractions to it. To apply the method, the sample has been further divided into three redshift 
bins to approximatively preserve the same number of 
galaxies in each bin, as shown in the figure. This choice is a trade-off aimed to maximize both the number of $H(z)$ measurements and to minimize the relative 
errorbars. The $D4000_{n}$ data have been then averaged in the previously discussed redshift bins using a weighted mean, 
to take into account also the error associated to each sample, since some points have a much smaller error than others; the results are shown in the upper plot of Fig. \ref{fig:Hz}.   
In the figure it is also shown the median $D4000_{n}$-z relation obtained in \cite{moresco2012a} in the same mass range ($10^{11}<M/M_{\odot}<10^{11.25}$), 
to give an insight of the overall evolution between $z\sim0$ to $z\sim2$. The last point of \cite{moresco2012a} will be also used as connection point between 
the low-redshift and the high-redshift analysis, to estimate the first $H(z)$ point.


\section {Method and Analysis}
\label{sec:analysis}

The Hubble parameter $H(z)$ is evaluated following eq. \ref{eq:HzD4000}, where the relevant quantities are $dz/dD4000_{n}$ and $A(SFH, Z/Z_{\odot})$.
To obtain the first one, it has been estimated the evolution between the last $D4000_{n}$ point provided in \citet{moresco2012a} and the first point of this analysis, 
and between the second and third points of this analysis. In this way, it was possible to provide two new $H(z)$ measurements.
The parameter $A(SFH, Z/Z_{\odot})$ has been estimated from stellar population synthesis models, as described in \citet{moresco2012a}. Briefly,
given the properties of the selected population, are considered models with an exponentially delayed SFH with different $\tau$ (up to $\tau$=0.3 Gyr, characteristic of 
the selected sample) and a stellar metallicity in the range $0.75<Z/Z_{\odot}<1.25$. This last choice is well justified both on observational 
and theoretical basis. On one hand, for the selected sample only broad constraints on metallicity are provided, with a general agreement on a near solar (or slightly 
subsolar) metallicity. The most precise constraint so far for this population of very massive and passive galaxies has been obtained by \citet{onodera2014}, finding 
from the stacked spectrum of 24 galaxies a metallicity in the range 0.024$\pm$ 20\%. On the other hand, these passive systems have already exhausted their 
gas reservoir, and therefore a negligible evolution is expected with respect to their low-redshift counterparts, which show an almost solar metallicity. Therefore, a 
conservative range of $\pm$25\% around solar metallicity is assumed.
Two stellar population synthesis models have been used: \cite{bruzual2003} (BC03) and \cite{maraston2011} (MaStro). They encompass different assumptions for stellar 
evolutionary models, treatment of the asymptotic giant branch phase, and method used to compute the integrated spectra, and so can be used 
to quantify the dependence of the results on the models. The parameter $A(SFH, Z/Z_{\odot})$ is estimated in both models, and the difference in the resulting
H(z) values has been summed in quadrature to the total error.

\begin{table}
\begin{center}
\caption{Hubble parameter measurements.}
\begin{tabular}{cc}
\hline \hline
z & $\rm H(z)$ [km/sec/Mpc]\\
\hline
1.363 & $160\pm33.6$\\
1.965 & $186.5\pm50.4$\\
\hline \hline
\end{tabular}
\label{tab:Hz}
\end{center}
\end{table}

In Tab. \ref{tab:Hz} and Fig. \ref{fig:Hz} are reported the two new H(z) constraints, and for comparison are shown also the values available in literature obtained 
from ``cosmic chronometers'' approach, both from relative ages or D4000. As it is possible to see, this analysis for the first time breaks the current limit at $z\sim1.75$, 
proving a measurement at $z\sim2$. These results have been obtained with just a few tens galaxies, and shows the potentiality of this approach to provide valuable 
constraints in the range $1.5<z<2$; in this range, forthcoming surveys such as Euclid \citep{euclid} and WFIRST \citep{wfirst}, will greatly improve the present-day 
statistic, increasing the number of passive galaxies at these redshifts by at least an order of magnitude \citep[see][]{euclid}.
In estimating the statistical strength of these new data, a simulation of the possible new constraints that can be achieved with new data was also performed.
It has been considered a Euclid-like survey observing approximatively one thousand spectra of passive galaxies in five redshift bins 
in the range $1.5<z<2$; the expected error is estimated taking into account, to be conservative, only the improvement due to the increased statistic. In this way, the systematic
errors are expected to clearly dominate over the statistical errors (see Fig. \ref{fig:Hz}). 
The data have been simulated on a Planck fiducial cosmology, shifting the simulated points from the theoretical relation with a gaussian distribution with dispersion 
equal to the estimated error. These simulated data are also shown in Fig. \ref{fig:Hz}.

\begin{figure*}
\includegraphics[angle=0, width=0.28\textwidth]{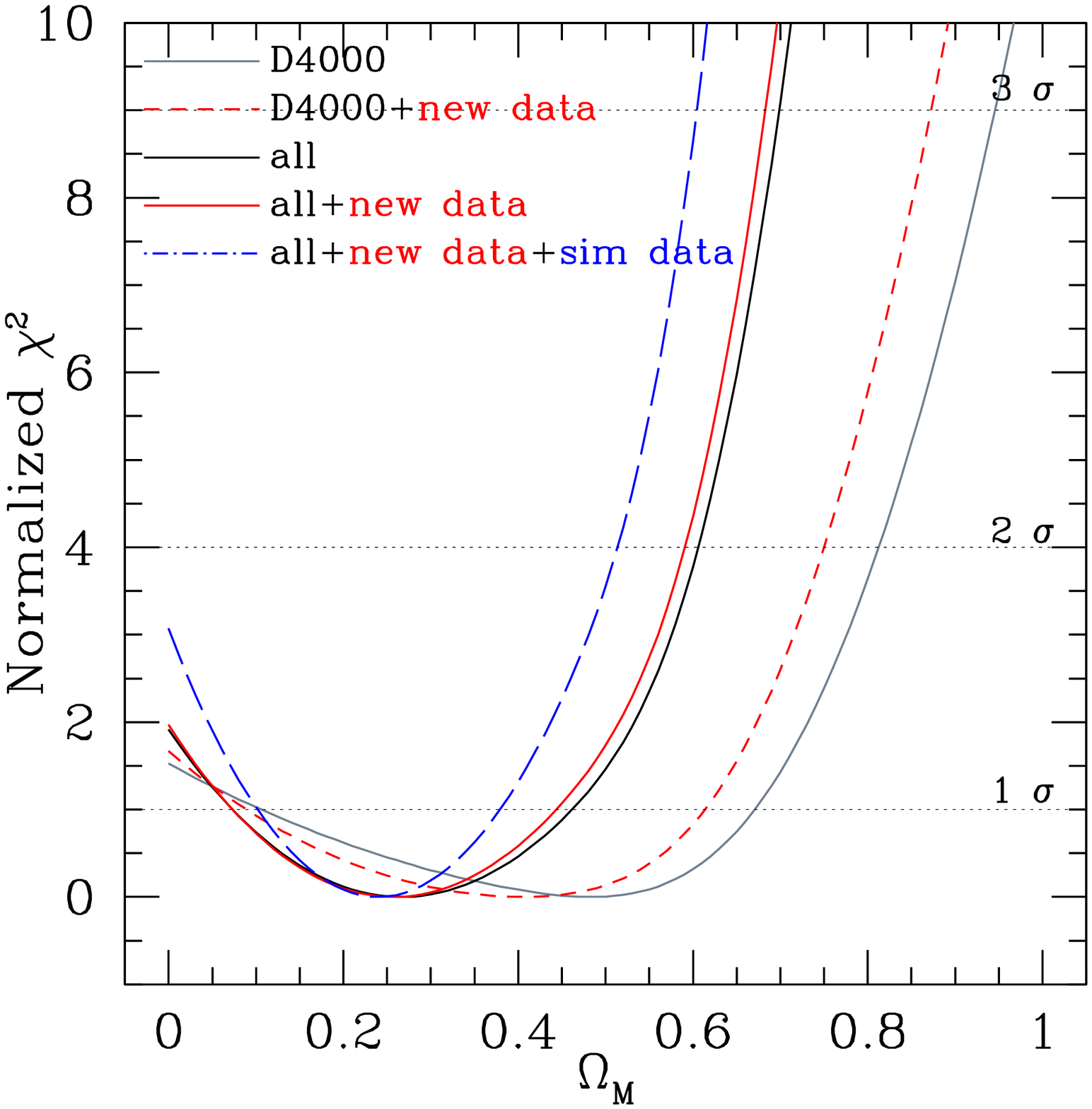}
\includegraphics[angle=0, width=0.28\textwidth]{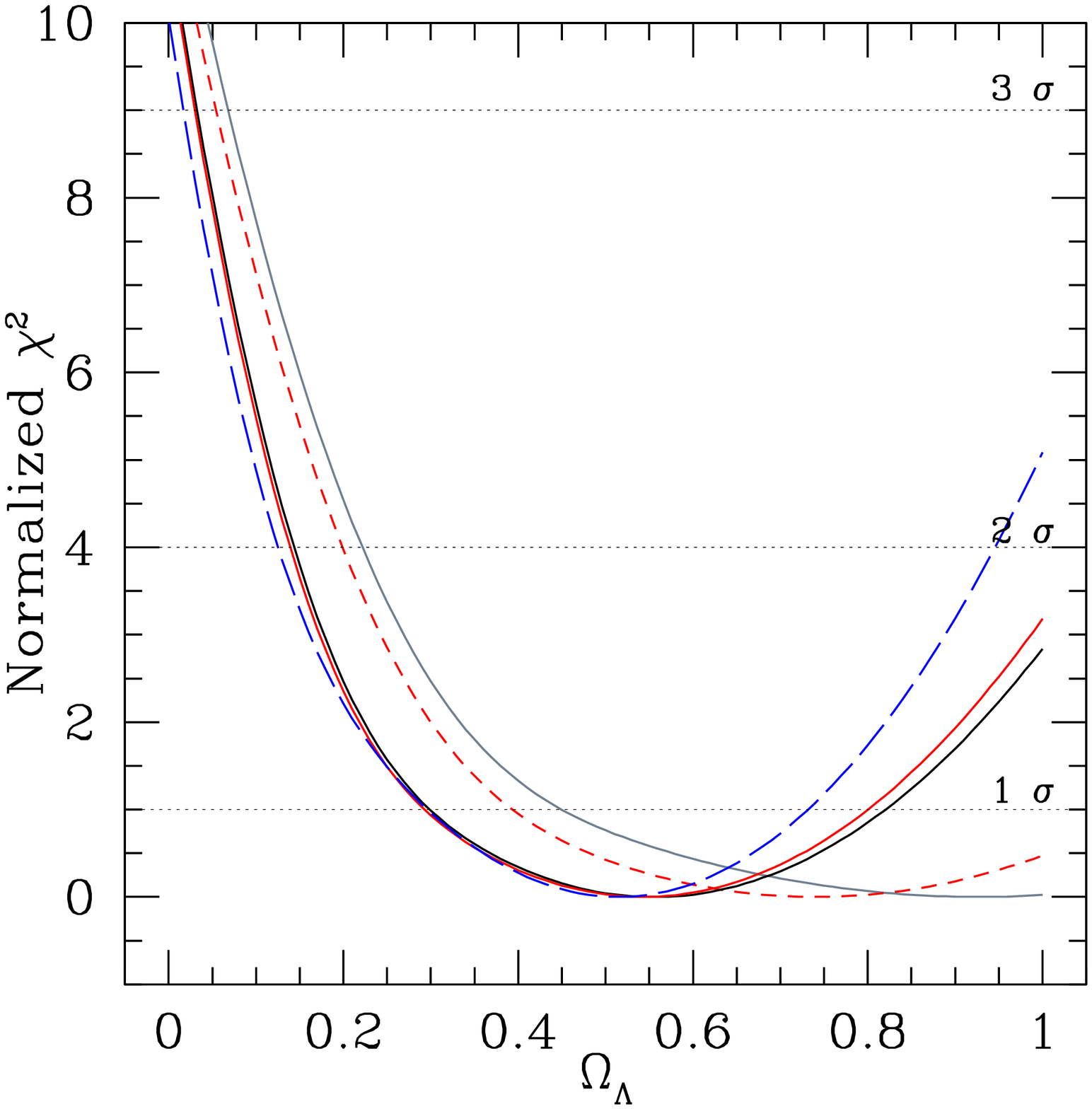}
\includegraphics[angle=0, width=0.28\textwidth]{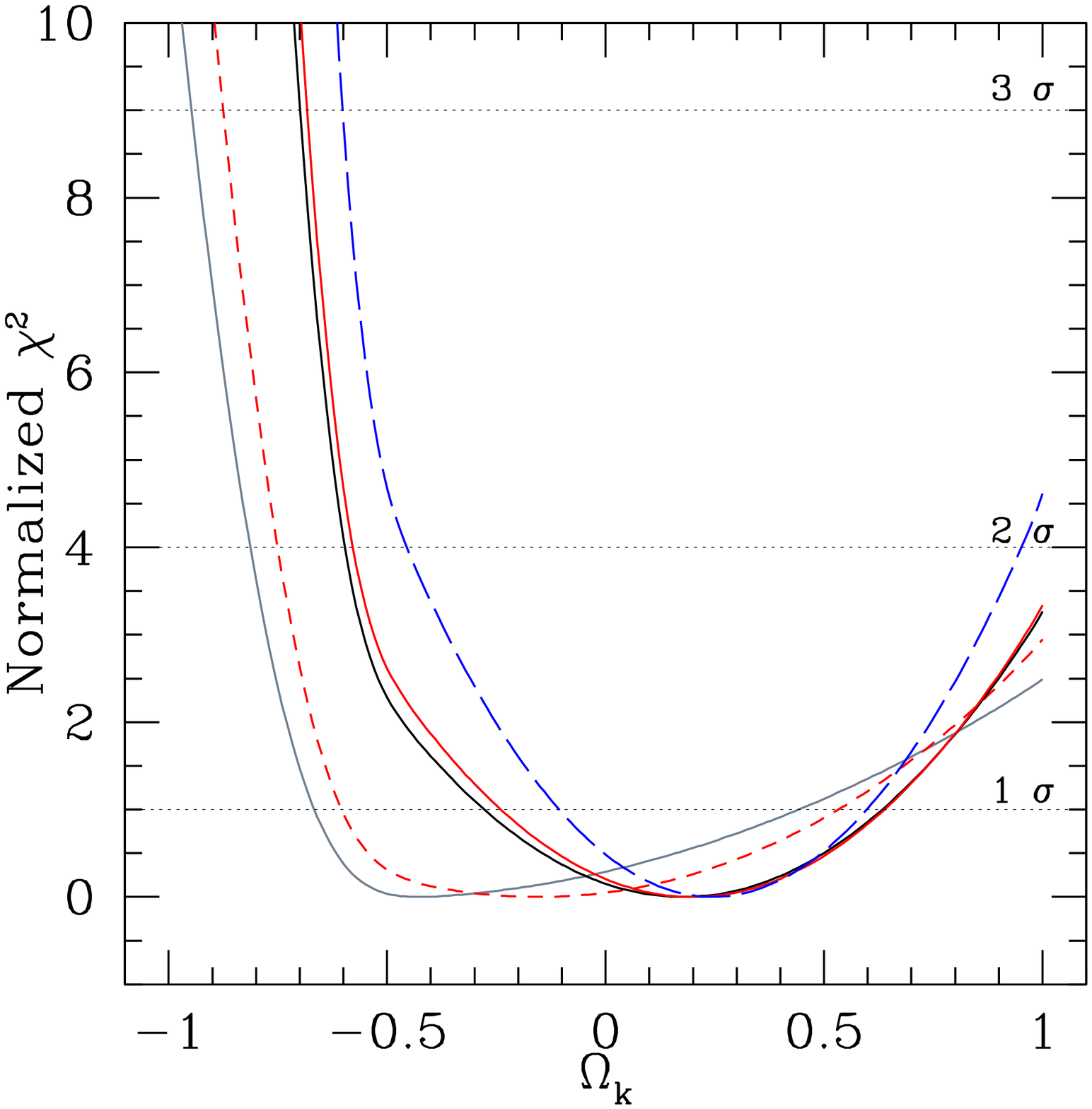}
\includegraphics[angle=0, width=0.28\textwidth]{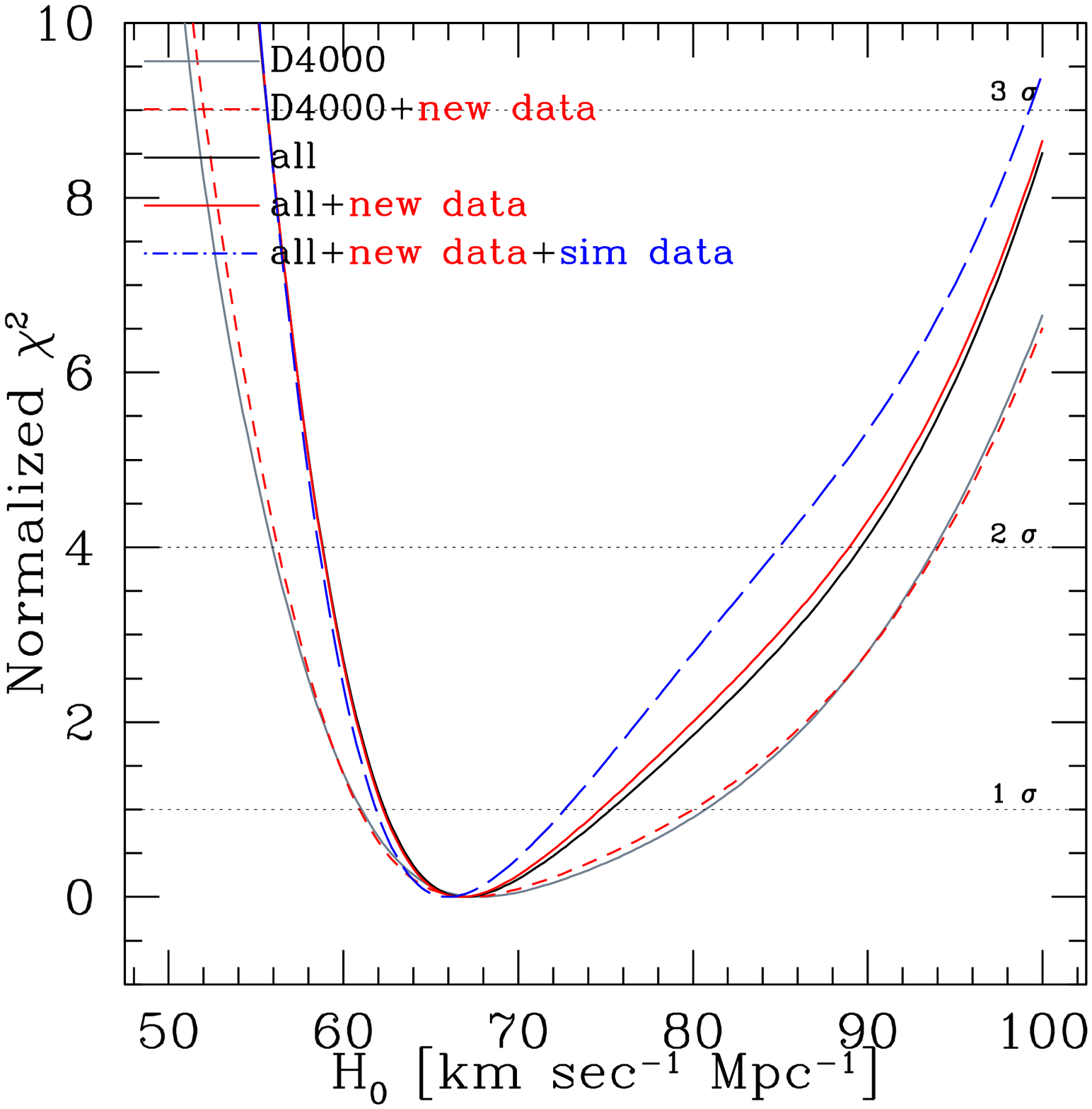}
\includegraphics[angle=0, width=0.28\textwidth]{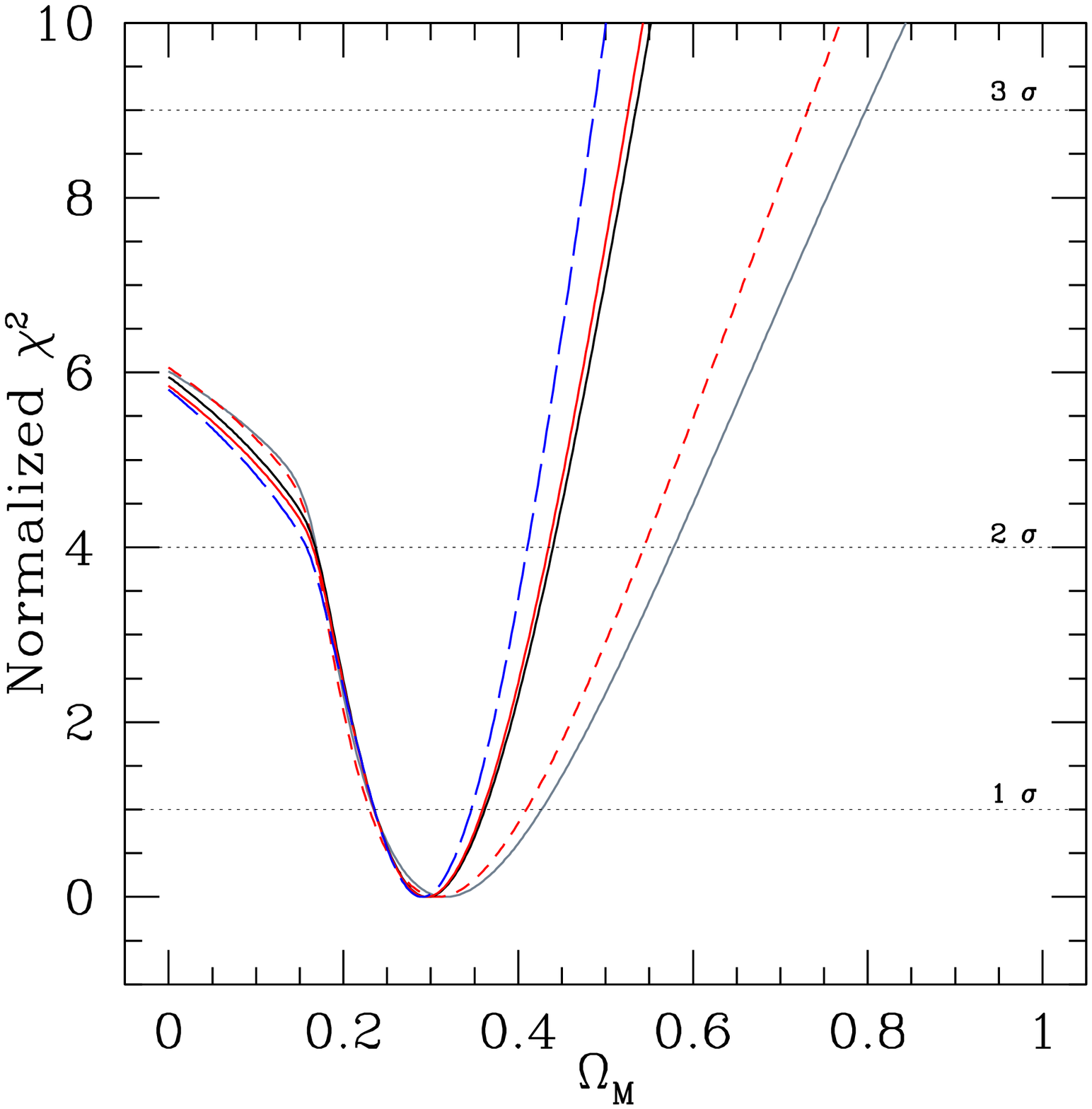}
\includegraphics[angle=0, width=0.28\textwidth]{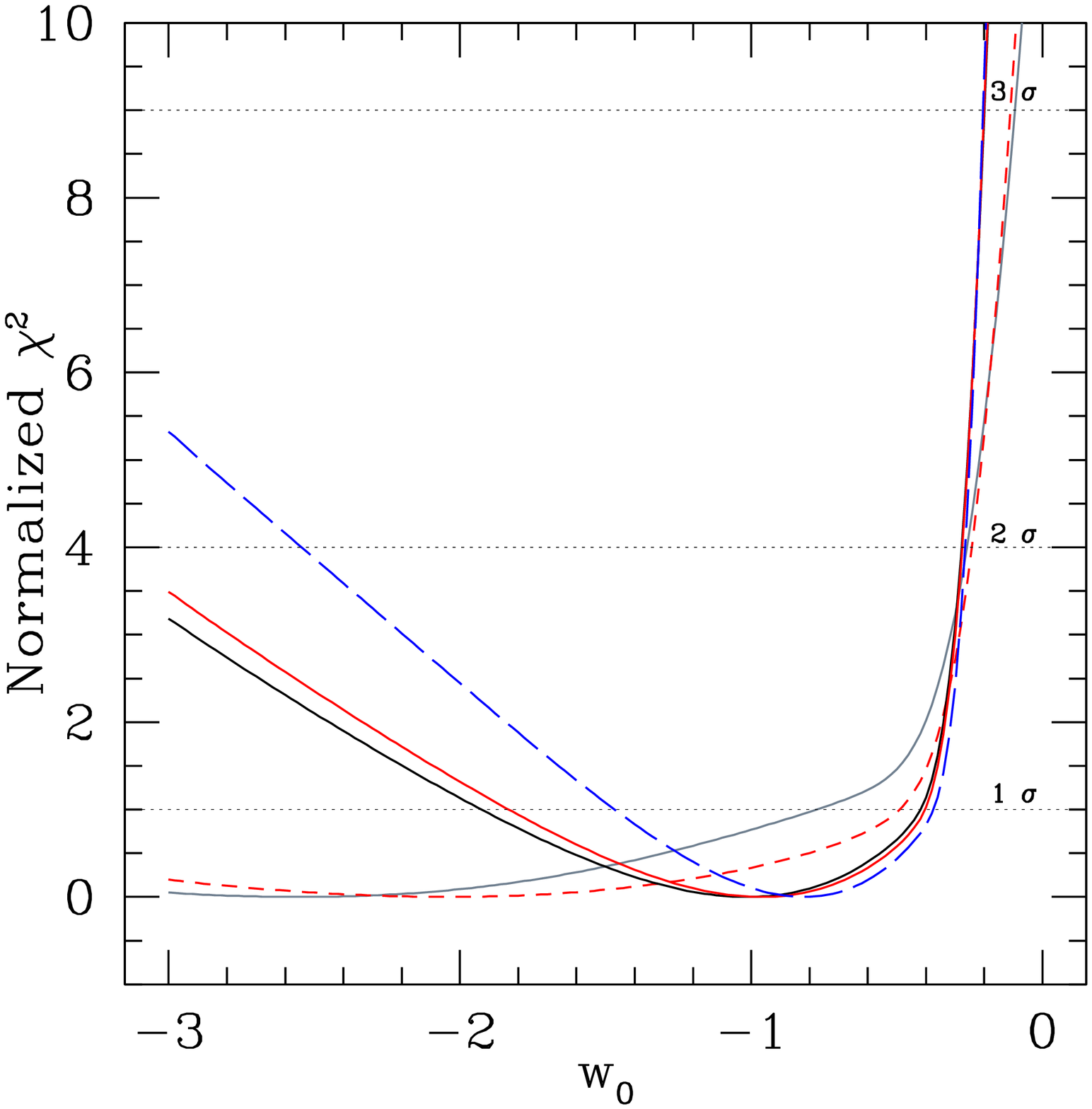}
\caption{One-dimensional marginalized $\chi^{2}$ for cosmological parameters in an open $\Lambda$CDM 
cosmology (upper panels) and in a flat wCDM cosmology (lower panels). In each case it is shown
for comparison the constraints obtained with D4000 data (with and without the new $H(z)$ measurements, grey
and red dashed lines), with all the data (with and without the new $H(z)$ measurements, black
and red solid lines), and also with the simulated data added (blue dashed lines).}
\label{fig:contours1D}
\end{figure*}

To explore the constraining power of the new H(z) data, I considered two models, an open $\Lambda$CDM model:
\begin{equation}
H(z)=H_{0}\sqrt{\Omega_{M}(1+z)^{3}+\Omega_{k}(1+z)^{2}+\Omega_{\Lambda}}
\end{equation}
where $\rm\Omega_{M}+\Omega_{\Lambda}+\Omega_{k}=1$ and a flat wCDM model:
\begin{equation}
H(z)=H_{0}\sqrt{\Omega_{M}(1+z)^{3}+\Omega_{\Lambda}(1+z)^{3(1+w_{0})}}
\end{equation}
A standard $\chi^{2}$ analysis is performed as a function of three cosmological parameters, i.e. 
$\rm\{\Omega_{M}, \Omega_{\Lambda}, \Omega_{k}\}$ \citep[H$_{0}$=67.3 km/sec/Mpc,][]{planck}
and $\rm\{\Omega_{M}, H_{0}, w_{0}\}$ respectively.
\begin{table}
\begin{center}
\caption{Marginalized cosmological parameters constraints obtained with cosmic chronometers in the two considered cosmologies.
The errors reported correspond to 1-$\sigma$ confidence levels.}
\begin{tabular}{lccc}
\hline \hline
o$\Lambda$CDM & $\Omega_{M}$ & $\Omega_{\Lambda}$ & $\Omega_{k}$\\
\hline
D4000 H(z) & $0.48^{+0.19}_{-0.37}$ & $0.93^{+0.07}_{-0.48}$ & $-0.43^{+0.87}_{-0.23}$\\
D4000 H(z) + new data & $0.41^{+0.2}_{-0.32}$ & $0.74^{+0.26}_{-0.34}$ & $-0.15^{+0.68}_{-0.45}$\\
all H(z) & $0.27^{+0.19}_{-0.19}$ & $0.55^{+0.27}_{-0.25}$ & $-0.19^{+0.44}_{-0.46}$\\
all H(z) + new data & $0.26^{+0.18}_{-0.18}$ &$0.54^{+0.26}_{-0.24}$ & $0.21^{+0.42}_{-0.45}$\\
all H(z) + new + sim data & $0.24^{+0.13}_{-0.13}$ & $0.51^{+0.22}_{-0.21}$ & $0.25^{+0.34}_{-0.35}$\\
\hline
fwCDM & H$_{0}$ [km/sec/Mpc]& $\Omega_{M}$ & $w_{0}$\\
\hline
D4000 H(z) & $67.8^{+12.8}_{-6.5}$ & $0.32^{+0.11}_{-0.08}$ & $<-0.78$\\
D4000 H(z) + new data & $67.3^{+12.5}_{-6.3}$ & $0.31^{+0.1}_{-0.08}$ & $<-0.5$\\
all H(z) & $67^{+8.3}_{-4.5}$ & $0.3^{+0.07}_{-0.06}$ & $-1^{+0.58}_{-0.92}$\\
all H(z) + new data & $66.8^{+7.8}_{-4.3}$ &$0.3^{+0.06}_{-0.06}$ & $-0.96^{+0.54}_{-0.86}$\\
all H(z) + new + sim data & $66^{+6.3}_{-4}$ & $0.29^{+0.06}_{-0.05}$ & $-0.82^{+0.44}_{-0.64}$\\
\hline \hline
\end{tabular}
\label{tab:constraints}
\end{center}
\end{table}
\begin{figure}
\includegraphics[angle=0, width=0.47\textwidth]{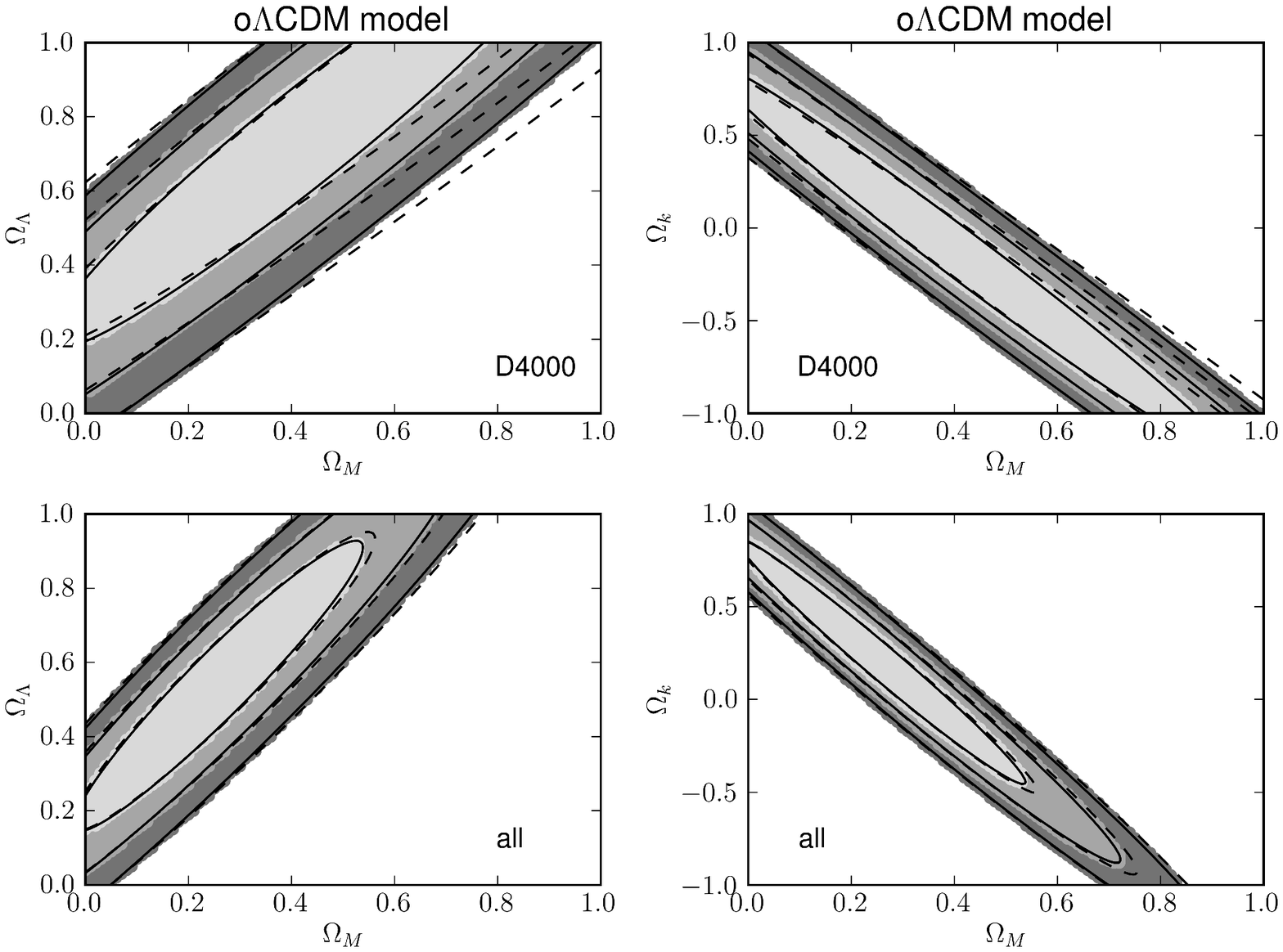}
\includegraphics[angle=0, width=0.47\textwidth]{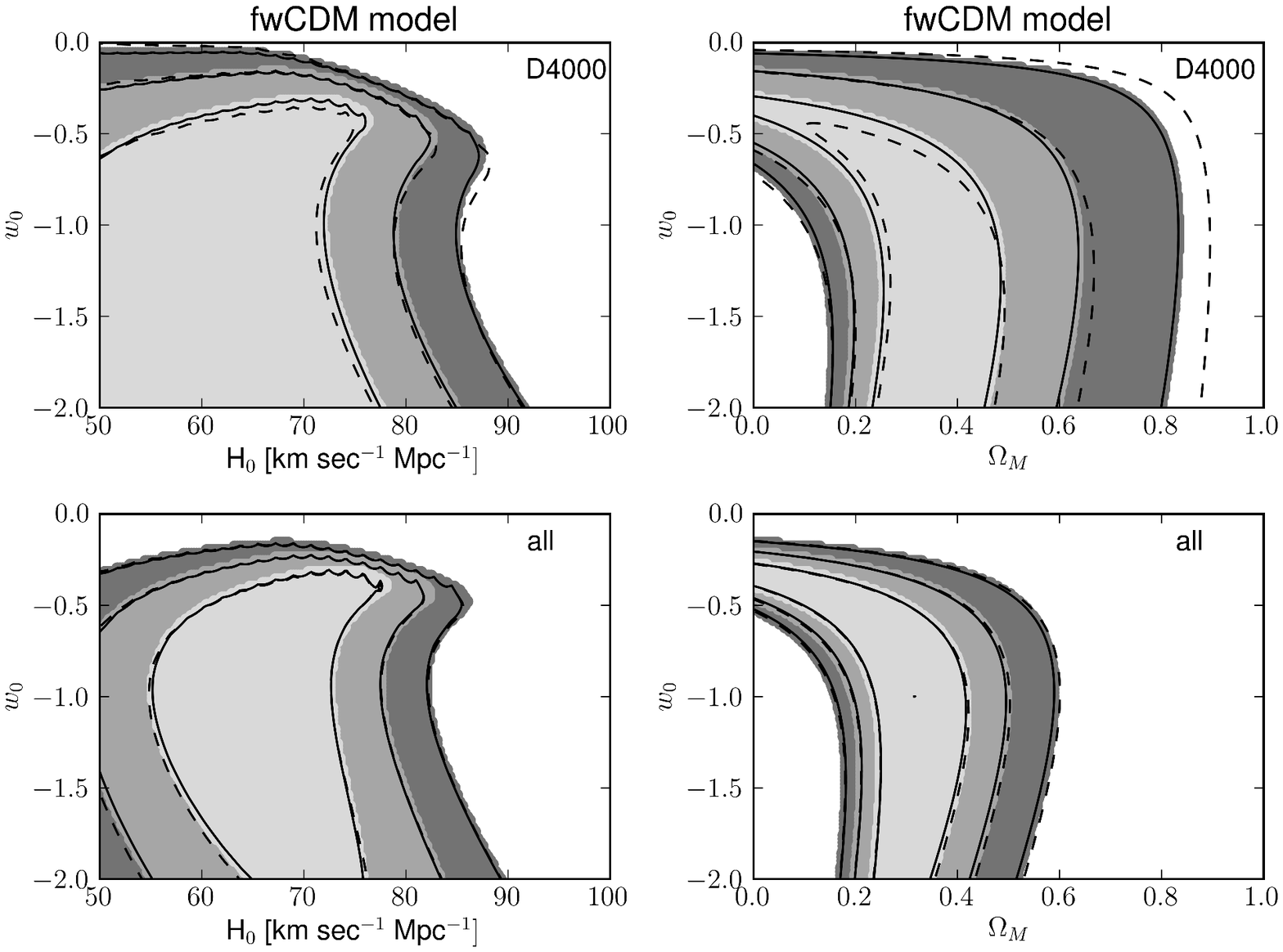}
\caption{Marginalized contour plots for an open $\Lambda$CDM 
cosmology (four upper panels) and in a flat wCDM cosmology (four lower panels).
In each cosmology are shown 1-$\sigma$, 2-$\sigma$, and 3-$\sigma$ confidence
levels (light, medium and dark gray respectively) considering only the D4000 data 
or all the data, as highlighted in the captions. Solid lines show the constraints obtained
including the new $H(z)$ data, while dashed lines show the constraints without them;
the relative gain in accuracy obtainable with these new data can be inferred by comparing 
the two lines.}
\label{fig:contours2D}
\end{figure}
The cosmological constraints obtained are shown in Tab. \ref{tab:constraints} and Figs. \ref{fig:contours2D} and
\ref{fig:contours1D}; the values provided have been obtained by marginalizing each time the full 3D likelihood 
on the uninteresting parameters. 
Both the constraints that can be achieved with only D4000 data and with all the H(z) measurements obtained from cosmic
chronometers are considered, and in both case the results obtained with and without the two new H(z) points of this analysis are 
compared. For many parameters, it is found a measurable, even if small, improvement adding the new
H(z) measurements, especially for $\Omega_{M}$ and $w_{0}$ ($\sim$5\%, see also Fig. \ref{fig:contours1D}); 
this is in particular remarkable considering that this analysis has been performed on a very small set of galaxies. 
The forecasts for a Euclid-like survey are more optimistic, with a percentage improvement on the error of $\sim$30\%, 
20\% and 15\% respectively for $\Omega_{M}$, $\Omega_{k}$ and $\Omega_{\Lambda}$ in a o$\Lambda$CDM 
cosmology, and $\sim$10\%, 15\% and 20\% respectively for $\Omega_{M}$, H$_{0}$ and w$_{0}$ in a flat wCDM 
cosmology.
As expected the improvement on the low-redshift parameter H$_{0}$ is less significant, since the new data 
primarily constrain the evolution of H(z) at high redshift.

\section{Conclusions} 
\label{sec:concl}
This analysis shows the potentiality of using the ``cosmic chronometers'' approach to constrain the expansion
history of the Universe up to $z\sim2$. Using spectroscopic observation of a few very massive and passive galaxies
in the range $1.4<z<2.2$, I provide two new measurements of H(z) at $z=1.36$ and $z=1.97$, improving the current
limit at $z\sim1.75$. The new data are shown to be important to further improve the accuracy especially $\Omega_{M}$ 
and $w_{0}$ with respect to current measurements obtained with this technique. The forecasts obtained on a Euclid-like 
survey confirmed an even higher improvement in the accuracy reachable on most cosmological parameters.
It is of particular interest that the provided constraints have been obtained with a single cosmological probe.
Much more precise estimates will be available once different probes are combined, as shown e.g. in \citet{moresco2012b}.\\


{\small {\it Acknowledgments.} I am grateful to A. Cimatti, L. Pozzetti, R. Jimenez and L. Verde for interesting 
comments and suggestions which helped to improve the presentation of this work.
I am also thankful to J.-K. Krogager and R. Gobat for providing detailed information 
about their samples, and to the anonymous referee for useful comments. 
I acknowledge financial contributions by grants ASI/INAF
I/023/12/0 ans PRIN MIUR 2010-2011 ``The dark Universe and the cosmic
evolution of baryons: from current surveys to Euclid''.}

\bibliographystyle{mn2e} \bibliography{bib}


\end{document}